\documentclass[final,5p,times,preprint]{elsarticle}
\usepackage[utf8]{inputenc}
\usepackage{amsfonts}
\usepackage{amsmath}
\usepackage{amssymb}
\usepackage{bm}
\usepackage{lineno,hyperref}
\usepackage{comment}
\usepackage{xcolor}
\usepackage{booktabs}
\modulolinenumbers[5]
\usepackage[normalem]{ulem}

\journal{Physics Letters B}

\biboptions{comma,sort&compress}

\begin{document}

\begin{frontmatter}

\title{Description of the $^{11}$Li$(p,d){^{10}}$Li transfer reaction using structure overlaps from a full three-body model}


\author[FAMN,ECT]{J. Casal\corref{mail}}
\cortext[mail]{Corresponding author}
\ead{casal@ectstar.eu}

\author[FAMN]{M. G\'omez-Ramos}

\author[FAMN]{A. M. Moro}

\address[FAMN]{Departamento de F\'{\i}sica At\'omica, Molecular y Nuclear, Facultad de F\'{\i}sica, Universidad de Sevilla, Apartado 1065, E-41080 Sevilla, Spain}
\address[ECT]{European Centre for Theoretical Studies in Nuclear Physics and Related Areas (ECT$^*$) and Fondazione Bruno Kessler, Villa Tambosi, Strada delle Tabarelle 286, I-38123 Villazzano (TN), Italy}

\begin{abstract}
Recent data on the differential angular distribution for the transfer reaction $^{11}$Li(p,d)$^{10}$Li at $E/A=5.7$~MeV in inverse kinematics are analyzed within the DWBA reaction framework, using the overlap functions calculated within a three-body model of $^{11}$Li. The weight of the different $^{10}$Li configurations in the system's ground state is obtained from the structure calculations unambiguously. The effect of the $^{9}$Li spin in the calculated observables is also investigated. We find that, although all the considered models succeed in reproducing the shape of the data, the magnitude is very sensitive to the content of  $p_{1/2}$ wave in the $^{11}$Li ground-state wave function. Among the considered models, the best agreement with the data is obtained when the $^{11}$Li ground state contains a $\sim$31\% of $p_{1/2}$ wave in the $n$-$^9$Li subsystem. Although this model takes into account explicitly the splitting of the $1^+$ and $2^+$ resonances due to the coupling of the $p_{1/2}$ wave to the $3/2^-$ spin of the core, a similar degree of agreement can be achieved with a model in which the $^{9}$Li spin is ignored, provided that it contains a similar $p$-wave content.
\end{abstract}

\begin{keyword}
$^{10,11}$Li\sep Transfer\sep DWBA \sep Overlaps \sep Three-body
\end{keyword}

\end{frontmatter}


\section{\label{sec:intro} Introduction}
Halo nuclei have triggered intensive work in the nuclear physics community since their discovery back in the eighties~\cite{Tanihata85,Hansen87}. The case of neutron Borromean nuclei, consisting of a compact core plus two valence neutrons~\cite{Zhukov93}, keeps being the subject of a considerable amount of experimental and theoretical studies. In these three-body systems, any two-body pair is unbound, which poses a real challenge from the theoretical point of view~\cite{Nielsen01}. Examples of two-neutron halo nuclei are $^{6}$He, $^{11}$Li, $^{14}$Be or $^{22}$C. The understanding of their structure requires solid constraints on the neutron unbound binary subsystems, i.e. $^5$He, $^{10}$Li, $^{13}$Be or $^{21}$C.

Our present understanding of the peculiar properties of these exotic systems largely stems from the analysis of reactions in which these nuclei are part of the colliding systems or appear within some of the reaction products. In the former case, the experiments must be performed in inverse kinematics, and have therefore only become  possible since the development of rare isotope beam facilities in the late eighties. Examples of these reactions are nucleon-removal (also named knockout) reactions \cite{Simon2007,Randisi2014}, Coulomb dissociation \cite{Aumann2013}, single- and multi-particle transfer and, most recently, quasi-free breakup reactions of the form (p,pn) or (p,2p) \cite{Kobayashi1997,Aksyutina2008,Kondo2010,Aksyutina2013}.   The outcomes of these measurements are complementary to one another.

Theoretical works have shown that the structure of two-neutron halo nuclei results from the delicate interplay  of several factors, such as the binding effect due to the pairing interaction between the halo neutrons, the coupling to the core collective excitations \cite{Barranco01}, the effect of Pauli blocking and the role of  tensor correlations \cite{Myo07}. 

In the case of $^{11}$Li, many theoretical and experimental efforts have been devoted to understanding its conspicuous structure. This system is bound by only $S_{2n}=369.15(65)$ keV \cite{Smith08}.  Its large spatial extension was first evidenced in the pioneering knockout experiments performed by Tanihata and collaborators \cite{Tanihata85} using an energetic $^{11}$Li beam on a $^{12}$C target.  The analysis of the momentum distributions from subsequent fragmentation experiments \cite{Nil95,Hum95,Zin95}, including the angular correlations of the fragments \cite{Simon99},  revealed an admixture of  $s$ and $p$ waves in the $^{11}$Li ground-state and permitted to extract their relative weights.  
Reaction cross section data of these high-energy experiments have been also used to constrain the radius and $s$-wave content of the $^{11}$Li ground state \cite{Alk96,Fortune15}.

 A proper understanding of the $^{11}$Li structure requires an accurate knowledge of the $^{10}$Li subsystem. This is an interesting system by itself,  located just beyond the drip-line for $N=7$ isotones, and displaying an inversion of the $2s_{1/2}$ and $1p_{1/2}$ levels. The location and properties of these levels has been studied by means of several knockout and transfer experiments (see \cite{Sanetullaev16} and references therein). Most of these experiments predict a near-threshold concentration of $\ell=0$ strength, consistent with the presence of a virtual state, and one or more narrow $\ell=1$ resonances with a centroid around $\sim$0.5~MeV. However, the detailed parameters for the virtual state (e.g.~scattering length) and the actual position and width of these resonances vary widely from one experiment to another and are still a matter of debate \cite{Fortune16}. The relative amount of $s$- and $p$-waves in the $^{11}$Li ground state varies also significantly between different works. Furthermore, none of these experiments was able to resolve the doublets arising from the coupling of the $s$-wave and $p$-wave to the $3/2^-$ spin of the $^9$Li  core. This introduces an additional ambiguity in the theoretical interpretation of these data, since the observed energy distributions are usually compatible with different model assumptions. 

Many of these experiments provide information on the continuum spectrum of the isolated $^{10}$Li, but not on the structure of $^{10}$Li within the $^{11}$Li nucleus.  
 One of the few exceptions is provided by a recent measurement of the $^{11}$Li(p,d)$^{10}$Li stripping reaction, performed at TRIUMF at $E/A=5.7$~MeV \cite{Sanetullaev16}. The measured excitation spectrum of the produced $^{10}$Li exhibits a  resonant-like structure located at $E_r=0.62(4)$~MeV. The comparison of the angular distribution of this peak, after subtraction of the non-resonant background, with standard DWBA calculations, was consistent with an $\ell=1$ configuration with a spectroscopic factor of $0.67\pm0.12$. This corresponds to a 33\% content of $(p_{1/2})^2$ in the $^{11}$Li ground state. In addition to the assumptions inherent to the DWBA reaction framework, the theoretical analysis was  based on a simple model for the overlap function of the transferred neutron, which was described using a Woods-Saxon well with standard parameters. While this model can be roughly justified for two-body systems, the underlying picture of a neutron orbiting a $^{10}$Li {\it core} seems more questionable. Consequently, some of the conclusions of Ref.~\cite{Sanetullaev16} could be biased by this oversimplification of the structure model. 

In the present work, we reexamine the same data using also the DWBA method, but replacing the simple Woods-Saxon model by a more sophisticated and, in principle, realistic description of the $^{11}$Li and $^{10}$Li systems. The former is treated within a three-body model, with effective $n$-$n$ and $n$+core interactions, whereas the $^{10}$Li is described using $n$+$^{9}$Li scattering states generated with the  same n-core interaction as that used for $^{11}$Li. Our goal is to clarify the influence of the structure model on the extracted properties and, in particular, to see whether the conclusions of~\cite{Sanetullaev16} are affected by the use of a more realistic structure model.



\section{\label{sec:formal} Reaction Formalism}

\begin{figure}[t]
\centering
\includegraphics[width=0.7\linewidth]{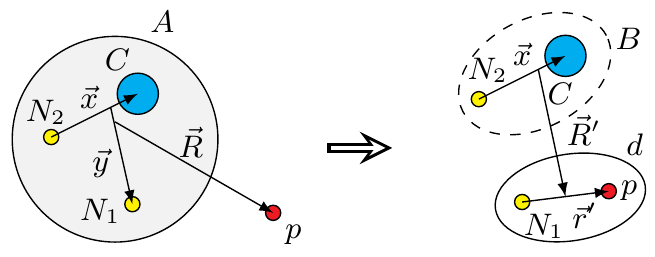}
\caption{Diagram for a $(p,d)$ reaction induced by a three-body projectile in inverse kinematics.}
\label{fig:scheme}
\end{figure}

Let us consider a reaction in which an incident composite nucleus $(A)$ collides with a proton target, which  picks one neutron $N_1$ from the projectile, giving rise to a residual subsystem $(B)$ and a deuteron ($d$).  We focus on the particular case of a three-body projectile comprising an inert core $(C)$ plus two valence neutrons $(N1, N2)$, so the reaction takes the form
\begin{equation}
\underbrace{(C+N_1+N_2)}_A + p \rightarrow \underbrace{(C+N_2)}_{B} + d,
\label{eq:scheme}
\end{equation}
which is schematically depicted in Fig.~\ref{fig:scheme}. If the nucleus $B$ does not form bound states (e.g., the composite $A$ is a Borromean system) the products of its decay after one neutron removal will provide spectroscopic information on the original projectile wave function. For this process, the prior-form transition amplitude can be formally reduced to an effective few-body problem, leading to
\begin{equation}
\mathcal{T}_{if}=\sqrt{2}\langle \Psi_{f}^{(-)}(\vec{x},\vec{R}',\vec{r}')|V_{pN_1}+U_{pB}-U_{pA} |\Phi_A(\vec{x},\vec{y}) \chi_{pA}^{(+)}(\vec{R}) \rangle,
\label{eq:tmatrix}
\end{equation}
where $\Phi_A$ represents the ground-state wave function of the initial three-body composite ($^{11}$Li in our case), 
 $\chi^{(+)}_{pA}$ is the distorted wave generated by the auxiliary potential $U_{pA}$, and $\Psi^{(-)}_f$ is the exact four-body wave function for the outgoing $d$-$B$ system. The $\pm$ superscript refers to the usual ingoing or outgoing boundary conditions. Notice the explicit factor $\sqrt{2}$ arising from the two identical neutrons in the initial wave function. The origin of this factor is further discussed in Ref.~\cite{GLENDENNING198345}. 

In writing the transition amplitude in the form (\ref{eq:tmatrix}), we implicitly use a {\it participant/spectator} approximation,  assuming that the  reaction occurs due to the interaction of the incident proton with a single neutron $(N_1)$ of $A$ (the {\it participant}), whereas the system $B=N_2+C$ remains unperturbed.

To reduce (\ref{eq:tmatrix}) to a tractable form, we approximate the exact wave function $\Psi^{(-)}_f$ by the factorized form,
\begin{equation}
\label{eq:dwba}
\Psi_{f}^{(-)}(\vec{x},\vec{R}',\vec{r}') \approx \varphi^{(-)}_{\vec{q},\sigma_2,\zeta}(\vec{x}) \chi_{dB}^{(-)}(\vec{R}')\phi_d(\vec{r}'),
\end{equation}
where $\phi_d$ is the deuteron wave function, $\varphi^{(-)}_{\vec{q},\sigma_2,\zeta}$ is a two-body continuum wave function with wave number $\vec{q}$ and spin projections ${\sigma_2,\zeta}$ of the binary subsystem $B$ ($^{10}$Li), and 
$\chi_{dB}^{(-)}$ is a distorted wave describing the $d$-$B$ relative motion in the exit channel. The function 
 $\varphi^{(-)}_{\vec{q},\sigma_2,\zeta}$ is the
time-reversed of $\varphi^{(+)}_{\vec{q},\sigma_2,\zeta}$, which can be written as~(c.f.~\cite{Satchler1983}, p.~135)
\begin{equation}
\begin{split}
\varphi^{(+)}_{\vec{q},\sigma_2,\zeta}(\vec{x}) &      = \frac{4\pi}{qx}\sum_{LJJ_TM_T}i^{L}Y^{*}_{LM}(\widehat{q}) 
						     \langle LMs_2\sigma_2|JM_J\rangle  \\
					    & \times \langle JM_JI\zeta|J_TM_T\rangle  f_{LJ}^{J_T}(qx) \left[\mathcal{Y}_{Ls_2J}(\widehat{x})\otimes\kappa_I\right]_{J_TM_T},
\end{split}
\label{eq:2bcont}
\end{equation}
where, for each component, the orbital angular momentum $\vec{L}$ and the spin $\vec{s}_2$ of the neutron $N_2$ couple to $\vec{J}$, and $\vec{J}_T$ results from coupling $\vec{J}$ with the spin $\vec{I}$ of the core. Note that $\vec{x}$ contains also the internal coordinates of $C$. The radial functions $f_{LJ}^{J_T}$ are obtained by direct integration of the two-body Schrodinger equation for the $N_2+C$ system subject to the boundary condition
\begin{equation}
f_{LJ}^{J_T}(qx) \longrightarrow \frac{i}{2}\left[H_L^{(-)}(qx) - S_{LJ}^{J_T} H_L^{(+)}(qx)\right], 
\label{eq:boundary}
\end{equation}
where  $q$ is related to the  $N_2+C$ relative energy as $q=\sqrt{2 \mu_x \varepsilon}/\hbar$, with $\mu_x$ its reduced mass, and $H^{(\pm)}$ are Coulomb functions~\cite{Thompson86}.

The ground state wave function of the initial three-body composite $A$ is described within a full three-body model~\cite{Descouvemont03,FaCE,MRoGa05}. In this work, this wave function is calculated as an expansion in hyperspherical harmonics using a pseudostate basis for the radial part \cite{JCasal13}, which has been successfully applied to describe the structure and reaction observables for exotic nuclei~\cite{JCasal14,JCasal15}.  This three-body wave function is most naturally obtained in the Jacobi-T set, but for the purposes of computing the required overlap functions, it is then transformed into the Jacoby-Y set, where the $x$ coordinate relates the core and one neutron (see Fig.~\ref{fig:scheme}), and we choose a coupling order compatible with that of the two-body continuum wave function given by Eq.~(\ref{eq:2bcont}). After diagonalization of the three-body Hamiltonian, the wave function can be written schematically as
\begin{equation}
\begin{split}
\Phi_A^{j\mu}(\vec{x},\vec{y}) & = \sum_{\beta_{3b}}w_{\beta_{3b}}^{j}(x,y) \\
& \times \left\{\left[\mathcal{Y}_{l_xs_2j_x}(\widehat{x})\otimes\kappa_{I}\right]_{j_1}\otimes\left[Y_{l_y}(\widehat{y})\otimes\kappa_{s_1}\right]_{j_2}\right\}_{j\mu},
\label{eq:3bwf}
\end{split}
\end{equation}
where $\beta_{3b}=\{K,l_x,j_x,j_1,l_y,j_2\}$ is a set of quantum numbers coupled to a total angular momentum $j$. In this set, $K$ is the hypermomentum, $l_x$ and $l_y$ are the orbital angular momenta associated with the Jacobi coordinates $x$ and $y$, respectively, and $j_x$ results from coupling $l_x$ with the spin $s_2$ of a single neutron.

Consistently with our spectator approximation for the composite $B$, we assume that the interaction does not change the state of this system, and the overlap between the two- and three-body wave functions (given by Eqs.~(\ref{eq:2bcont}) and (\ref{eq:3bwf}), respectively) contain all the relevant structure information. If we denote by $\psi_{LJJ_TM_T}$ the part of this overlap containing the spatial dependence in Eq.~(\ref{eq:2bcont}), we have
\begin{equation}
\psi_{LJJ_TM_T}(q,\vec{y}) = \int \frac{f_{LJ}^{J_T}(qx)}{x}\left[\mathcal{Y}_{Ls_2J}(\widehat{x})\otimes\kappa_I\right]^*_{J_TM_T}\Phi_A^{j\mu}(\vec{x},\vec{y})d\vec{x},
\label{eq:overlap1}
\end{equation}
and the transition amplitude $\mathcal{T}_{if}$ can be expanded as
\begin{equation}
\begin{split}
\mathcal{T}_{if}  & = \sqrt{2}\frac{4\pi}{q} \sum_{LJJ_TM_T} (-i)^{L} Y_{LM}(\widehat{q}) \langle LMs_2\sigma_2|JM_J\rangle \\ & \times\langle JM_JI\zeta|J_TM_T\rangle \mathcal{T}_{if}^{LJJ_TM_T},
\label{eq:tmatrix2}
\end{split}
\end{equation}
depending on a set of DWBA-like amplitudes,
\begin{equation}
\mathcal{T}_{if}^{LJJ_TM_T} \equiv \langle \chi_{dB}^{(-)}\phi_d|V_{pN_1}+U_{pB}-U_{pA} |\psi_{LJJ_TM_T}~\chi_{pA}^{(+)} \rangle.
\label{eq:tJ}
\end{equation}
These amplitudes enable the description of the process using a consistent structure input, in which the three-body projectile and the binary fragment incorporate the same core-neutron interaction.
From the transition amplitude, and after integrating over the angles $\widehat{q}$ of the relative wave vector $\vec{q}$, the double differential cross sections as a function of the $C$-$N_2$ relative energy and the scattering angle of $B$ with respect to the incident direction can be written as
\begin{align}
\frac{d\sigma^2}{d\Omega_B d\varepsilon_x} & = \frac{32\pi^2}{q^2}\rho(\varepsilon_x) \frac{1}{2(2j+1)}\frac{\mu_i\mu_f}{(2\pi\hbar^2)^2}\frac{k_f}{k_i} \nonumber \\
&\times\sum_{LJJ_T}\sum_{\nu}\left|\mathcal{T}_{if}^{LJJ_TM_T}\right|^2,
\label{eq:doubleTj}
\end{align}
where $\rho(\varepsilon_x)=\mu_x q/[(2\pi)^3 \hbar^2]$ is the density of $B$ states as a function of the $C$-$N_2$ excitation energy $\varepsilon_x$, $\mu_{i,f}$ the projectile-target reduced mass in the initial and final partitions, and $\nu\equiv\{M_T\sigma_d\}$ represents the spin projections of the final products.

\begin{table*}[t]
\centering
\begin{tabular}{cccccccc}
\toprule
Model  & $v_c^{(0)}$  &   $v_c^{(1)}$ &  $v_{ss}^{(0)}$  & $v_{ss}^{(1)}$  &    $v_\text{so-c}^{(1)}$ &     $v_\text{so-v}^{(1)}$ \\ 
\midrule
P1I &  -5.4          &   260.75           &    -4.50            &    1.00           &    1.00                  & 300                       \\
P2I &  -5.0          &   260.25           &    -2.00            &    1.00           &    1.00                  & 300                       \\
P3    &  -50.5         &   -39.0          &       --         &     --          &        --              &       40.0                \\
P4    &  -49.6         &   -39.4          &       --         &   --            &        --              &       35.5                \\
\bottomrule
\end{tabular}
\caption{Potential strengths for the $^{11}$Li models used in this work. Potentials P1I and P2I follow Eq.~(\ref{eq:p1}) and have a Gaussian radial shape with $a=2.55$~fm. Potentials P3 and P4 follow Eq.~(\ref{eq:p2}) and use Woods-Saxon forms with $R=2.642$~fm and $a=0.67$~fm.}
\label{tab:param}
\end{table*}

\section{\label{sec:models} Structure model}
To describe the $^{11}$Li system, we need the $n$-$n$ and $n$-$^{9}$Li interactions.  For the former, we use the well known GPT potential \cite{GPT}, which reproduces the scattering length for this system. For $n$-$^{9}$Li, since the properties of $^{10}$Li are less known, we consider four different choices, which will produce four different ground state wave functions of $^{11}$Li. This will allow us to study the influence of the underlying properties of the $^{11}$Li nucleus on the calculated cross sections. The first two models, labeled  P1I and P2I hereafter, are based on the following parametrization of Garrido {\textit{et al.} \cite{Garrido03}, with parity-dependent central, spin-orbit and spin-spin components
\begin{equation}
\label{eq:p1}
V_{^9\text{Li-}n}^{(L)}(x) = V_c^{(L)}(x) + V_{ss}^{(L)}(x) \vec{s}_2\cdot\vec{I} + V_\text{so-c}^{(L)}(x) \vec{L}\cdot\vec{I} + V_\text{so-v}^{(L)}\vec{L}\cdot\vec{s}_2.
\end{equation}
These models take into account the actual $^{9}$Li spin ($I=3/2^-$), which results in a splitting of the single particle orbital $s_{1/2}$ into $1^-,2^-$ states, and $p_{1/2}$ into $1^+,2^+$ resonances. The radial functions follow the Gaussian shape $V_i^{(L)}(x) = v_i^{(L)} \exp\{-(x/R)^2\}$, with $R=2.55$ fm and the strength parameters listed in Table~\ref{tab:param}. 

The other two models, labeled P3 and P4, are based on the parametrization of Thompson and Zhukov \cite{Thompson94}. They ignore the spin of the $^{9}$Li core and contain central and spin-orbit terms,
\begin{equation}
\label{eq:p2}
V_{^9\text{Li-}n}^{(L)}(x) = V_c^{(L)}(x) + V_\text{so-v}^{(L)}\vec{L}\cdot\vec{s}_2,
\end{equation}
where the central term follows the usual Woods-Saxon form $V_c^{(L)}(x) = v_c^{(L)}/(1+ \exp\{(x-R)/a\})$, the spin-orbit is given by a Woods-Saxon derivative, and their parameters are $R=2.642$~fm and $a=0.67$~fm. The strengths are also shown in Table~\ref{tab:param}. 

With these potentials, the ground state of $^{10}$Li is a $s_{1/2}$ virtual state. In the models including the spin of the core, the strengths of the different terms are chosen to ensure that, after the spin-spin splitting, the $2^-$ state is the ground state of the system and the 2$^+$ resonance appears higher in energy than the 1$^+$ resonance. For each choice of the binary interactions, we compute the ground state of $^{11}$Li in our full three-body model. Potentials P1I and P2I include repulsive terms to suppress the $p_{3/2}$ Pauli forbidden states, which is referred to in the literature as the repulsive core approach~\cite{Thompson00}. On the contrary, P3 and P4 produce a $p_{3/2}$ bound state that needs to be removed in order to avoid unphysical states in the three-body calculations. Here, we use the adiabatic projection method~\cite{Garrido97}. Following previous works, to recover the ground-state energy we include also a three-body force to account for effects not explicitly included~\cite{JCasal13,RdDiego14}, such as pairing correlations or core excitations. For these calculations, we adopt the ground state energy $-0.37$~MeV~\cite{Smith08}.


The resulting properties of the $^{10}$Li and $^{11}$Li systems obtained with these four parametrizations are shown in Table~\ref{tab:pot}. 
For P1I, the 2$^-$ ground state of $^{10}$Li is characterized by a scattering length of $a_s=-37.9$ fm, and the $1^-$ states correspond to non-resonant continuum. For the $p_{1/2}$ resonances, this model provides two states at 0.37 and 0.61 MeV with spin and parity of $1^+$ and $2^+$, respectively. The three-body wave function probabilities in the $n$-$^9$Li subsystem are given by 31\% of $p_{1/2}$ components and 67\% of $s_{1/2}$ components. In this model, the $d_{5/2}$ contribution is negligible. The weights of the individual $1^-,2^-,1^+,2^+$ configurations are 27\%, 40\%, 12\% and 19\%, respectively. Taking these values into consideration, the effective scattering length of the 2$^-$ ground state is reduced to $a_\text{eff}=-29.3$ fm, and the two $p$ resonances have their centroid at 0.52 MeV. This is consistent with recent experiments being unable to resolve the doublet. The three-body calculations using this $n$-$^9$Li potential provide a $^{11}$Li ground state characterized by matter and charge radii of 3.2 fm and 2.41 fm, respectively. As we will see below, this potential including the $^9$Li spin is found to give the best agreement with experimental data.

The other potentials give different properties for $^{10}$Li and hence for $^{11}$Li. The potential P2I is chosen to produce two $s_{1/2}$ virtual states with smaller scattering lengths in absolute value and the two $p_{1/2}$ resonances at slightly lower energies. This enhances the $p_{1/2}$ content in the $^{11}$Li ground state up to 44\%, and reduces its radius by about 10\%. Regarding the interactions neglecting the spin of the core, P3 is chosen to give a similar $p_{1/2}$ content as P1I, while in P4 the $s_{1/2}$ content is strongly reduced. This is reflected by the different values of the scattering lengths shown in Table~\ref{tab:pot}. Note that P3 and P4 predict a small but not negligible $d$-wave component, so the $s_{1/2}$ and $p_{1/2}$ contributions add up to $\sim$94\% only. Notice also that P3 provides the largest matter distribution for $^{11}$Li. In the following, the effect of these structure properties on the reaction dynamics will be discussed.

\section{\label{sec:results} Application to $^{11}$Li(p,d)$^{10}$Li}

\begin{figure}[t]
\centering
\includegraphics[width=0.8\linewidth]{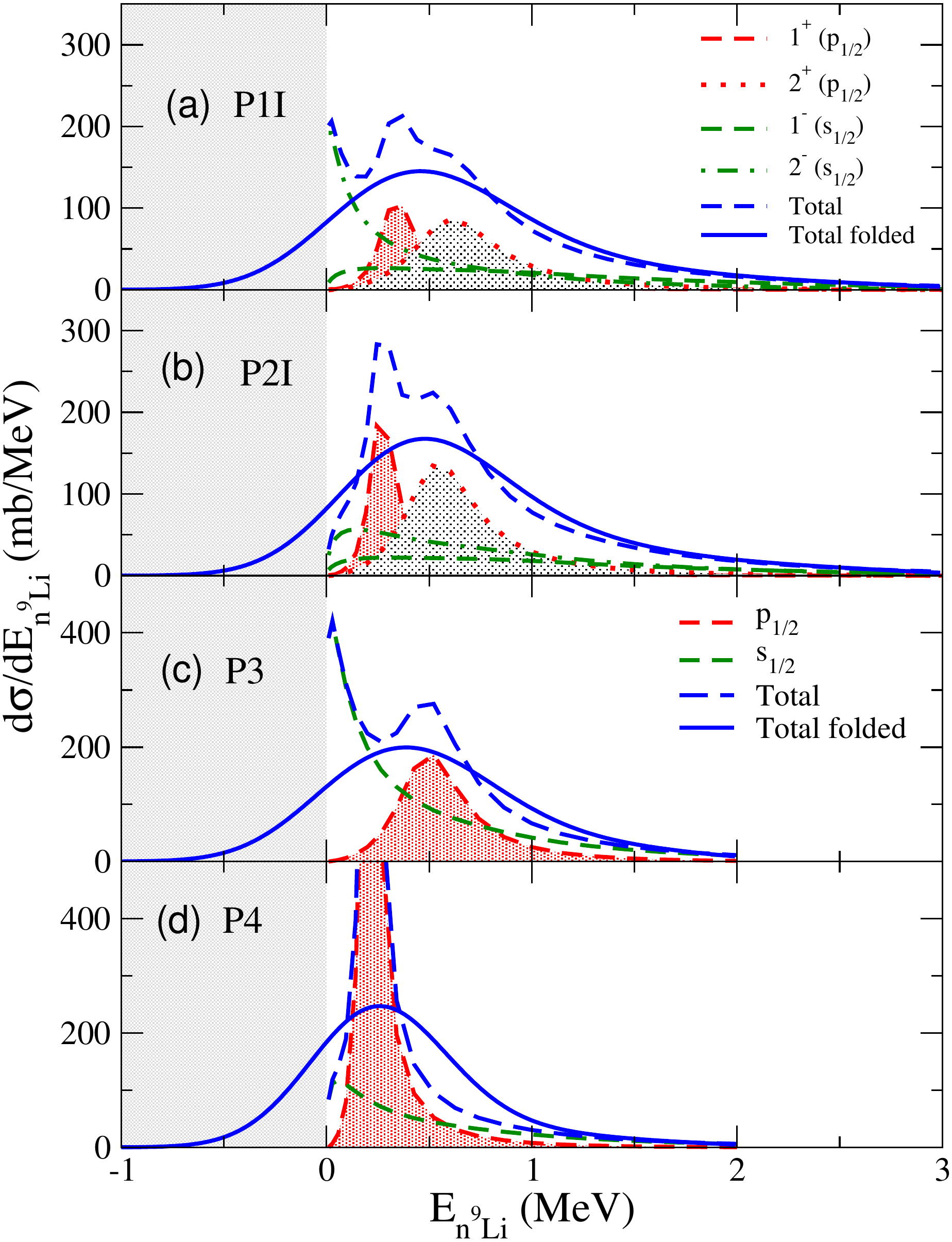}
\caption{The relative $n$-$^{9}$Li energy spectrum for the different $s_{1/2}$ (green) and $p_{1/2}$ contributions (red). The angle-integrated  relative-energy cross section before (blue dashed) and after (blue solid) the convolution with the experimental resolution are also shown. Potentials P1I, P3 and P4 have been selected for the calculations in the top, middle and bottom panels respectively. For P1I, the $1^-$ (green dashed) and $2^-$ (green dotted) contributions are shown separately as well as $1^+$ (red dashed) and $2^+$ (red dotted).}
\label{fig:dsde}
\end{figure}

\begin{figure}[t]
\centering
\includegraphics[width=1.0\linewidth]{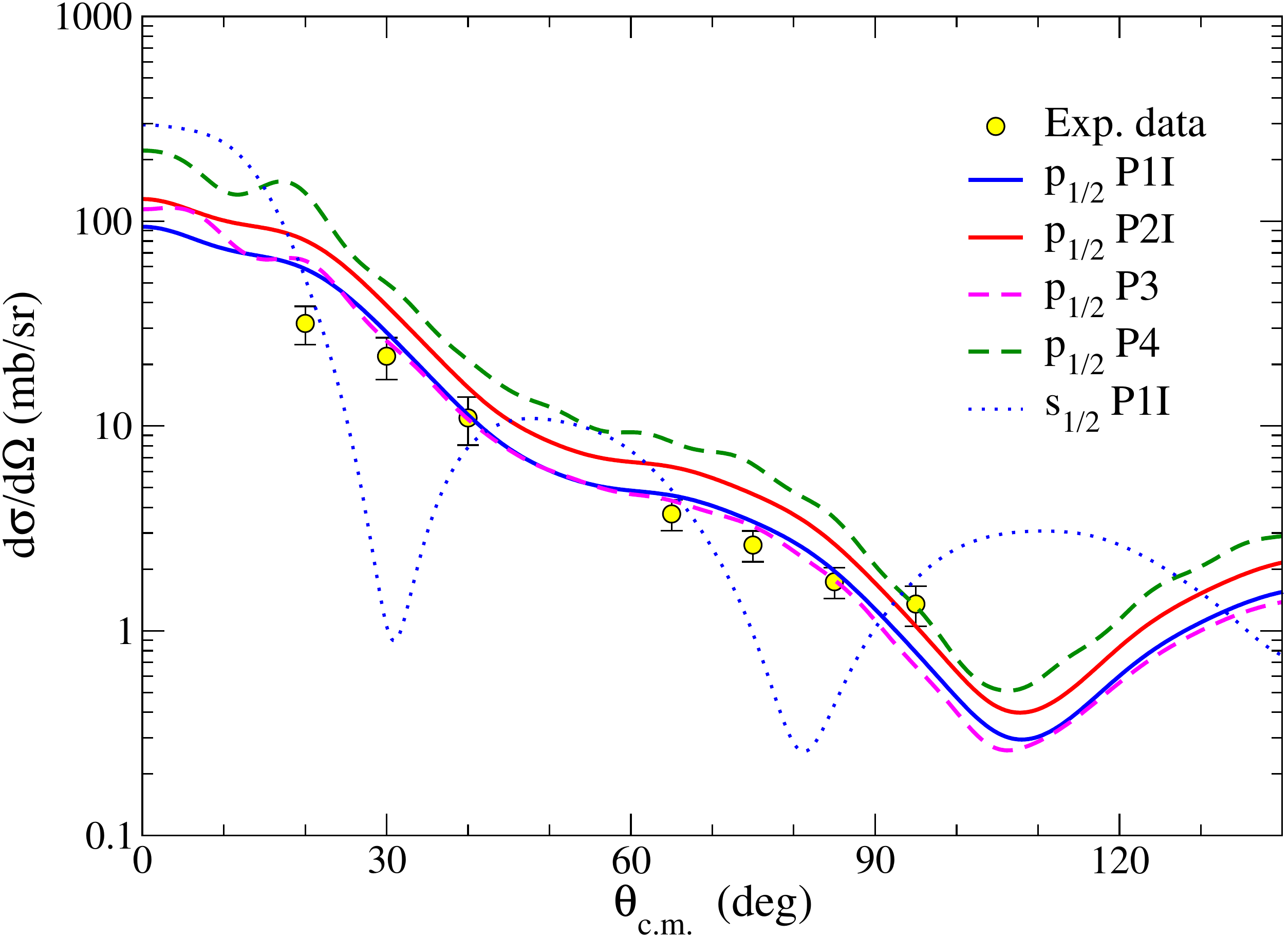}
\caption{Energy-integrated angular differential cross section for the $^{11}$Li$(p,d){^{10}}$Li reaction at 5.7 MeV/u. The lines are the results of DWBA calculations populating the $p_{1/2}$ orbitals in $^{10}$Li, using four different $^{11}$Li models: 
P1I (blue solid), P2I (red solid), P3 (magenta dashed) and P4 (green dashed). The cross section for neutrons transferred from the $s_{1/2}$ orbitals is also shown, for P1I, in the blue dotted line. Experimental data are from Ref.~\cite{Sanetullaev16}.}
\label{fig:dsdw}
\end{figure}

We apply the present formalism to the  reaction $^{11}$Li$(p,d){^{10}}$Li and compare our results with the recent experimental data by Sanetullaev \textit{et al.}~\cite{Sanetullaev16}.
As in Ref.~\cite{Sanetullaev16}, we describe the reaction mechanism in DWBA and use the same potentials for the $p\text{-}^{11}$Li, $d\text{-}^{10}$Li and $p\text{-}n$ interactions as in 
Ref.~\cite{Sanetullaev16}.  Therefore, the main difference between our calculations and those of \cite{Sanetullaev16} relies on the choice for the $\psi_{LJJ_TM_T}(q,\vec{y})$ overlaps. Whereas in \cite{Sanetullaev16} they were approximated by a single-particle wave function of the transferred  neutron in a Woods-Saxon potential, we obtain these functions from the three-body model of $^{11}$Li described in the preceding section. We must also remark that, while the calculation presented in \cite{Sanetullaev16} assumes a definite energy for the $^{10}$Li resonance, the calculations in this paper are performed for a range of $n\text{-}^{9}$Li continuum energies extending from 0 to 3 MeV, and integrations over energy have been performed when necessary.

\begin{table*}[t]
\centering
\begin{tabular}{cccccccccc}
\toprule
   & \multicolumn{2}{c}{$E_r$ (MeV)}   &   \multicolumn{2}{c}{$a$ (fm)} & & \%$p_{1/2}$&\%$s_{1/2}$& $r_{mat}$ (fm) & $r_{ch}$ (fm)\\
   & $1^+$  & $2^+$   		    & 	$1^-$ & $2^-$  		    &  &            &           &           &         \\  
\toprule
P1I & 0.37 & 0.61  & --    & -37.9 & & 31 & 67 & 3.2  & 2.41\\
P2I & 0.30 & 0.55  & -1.1  & -6.7  & & 44 & 54 & 3.0  & 2.40\\
\midrule
P3 &  \multicolumn{2}{c}{0.50} &  \multicolumn{2}{c}{-29.8} & & 30 & 64 & 3.6 & 2.48\\
P4 &  \multicolumn{2}{c}{0.23} &  \multicolumn{2}{c}{-16.2} & & 67 & 27 & 3.3 & 2.43\\
\bottomrule
\end{tabular}
\caption{Features of the $^{10}$Li structure for the different potentials employed in this work. The second column shows the energy of the $p_{1/2}$ resonance while the third one shows the scattering length of the $s_{1/2}$ virtual state. Note that for the model with spin, both the resonance and virtual state are split. The fourth and fifth columns show the weights of the $p_{1/2}$ and $s_{1/2}$ waves in the $^{11}$Li ground state respectively, while the last two columns show its matter and charge radii.}
\label{tab:pot}
\end{table*}

In Fig.~\ref{fig:dsde}, the computed $n\text{-}^{9}$Li relative energy spectra are presented using the preceding sets of potential parameters for the $n\text{-} ^9\text{Li}$ interaction. In these calculations, it has been assumed that the two halo neutrons of $^{11}$Li have the same angular momentum $\vec{J}= \vec{L} + \vec{s}_2$. This condition is strictly fulfilled when the spin of the core is neglected and is also a good approximation when using $I=3/2^-$. Only the configurations $(p_{1/2})^2$ and $(s_{1/2})^2$ have been considered, since other contributions, while present in the three-body calculations, give negligible components in the ground state of $^{11}$Li. The four panels in Fig.~\ref{fig:dsde} correspond to the results using potentials P1I, P2I, P3 and P4.
Here, the shaded peaks correspond to the $p_{1/2}$ resonances. The splitting of the resonances and virtual states can be seen in the upper two spectra. A folding of the total distribution with the experimental resolution given in Ref.~\cite{Sanetullaev16} is presented for all four calculations (solid line). It is rather remarkable how the low-energy peaks are smoothed out by the experimental resolution leading to a relatively featureless distribution, which could explain why the fine details of $^{10}$Li are not observed experimentally. The peak widths and heights, however, can be noticeably different. In general, the effect of the spin-spin splitting broadens the energy distributions.

In Fig.~\ref{fig:dsdw} the energy-integrated angular distribution of the outgoing deuteron is presented. Calculations have been performed using again the four discussed potential models. It has been assumed that the subtraction of the non-resonant background performed in the analysis of the data completely excludes the $s$-wave component, so experimental data are compared only with the cross section corresponding to neutrons transferred from the $p_{1/2}$ orbitals. Here we must remark that no fitting to the data has been performed, since the spectroscopic factor is given by the $p$-wave content in the three-body calculation. Our calculations provide absolute cross sections, in contrast to the theoretical curves in Ref.~\cite{Sanetullaev16} and previous works in which the cross sections are renormalized to fit the experimental data and extract spectroscopic factors. The results for P1I and P3 give the best agreement with the experimental data, while the shape of all four calculations is rather similar in spite of the different structure properties of the $^{11}$Li ground state. This seems to portray the energy-integrated angular distribution as an observable dependent mostly on the angular momentum of the extracted neutron and its weight in the ground-state of $^{11}$Li. Note that the experimental data have been measured at angles for which the $s_{1/2}$ contribution is minimal, as shown in Fig.~\ref{fig:dsdw}, where the $s$-wave contribution corresponding to the model P1I is presented. This may explain why in our energy distribution the contribution from $s$ waves is quite important at low energies, while experimentally there is no direct sign of this effect.



The $p$-wave content suggested by the present analysis ($\sim$30\%) is very close to the value extracted in the original analysis of the same data \cite{Sanetullaev16}. 
However, it is somewhat smaller than the  value of 45\% extracted from the analysis of the momentum distributions in the $^{11}$Li fragmentation  on carbon \cite{Simon99}. A reanalysis of the same data, along with the data from Ref.~\cite{Hum95}, performed by Garrido {\it et al} \cite{Garrido02} using a $^{11}$Li three-body wave function similar to our PI models within a participant-spectator reaction model, suggested a $p$-wave content of about $40\%$. Another {\it inert} core model proposed by Vinh Mau and Pacheco \cite{Vin96}, which uses a density-dependent pairing interaction between the halo neutrons, gave a $p_{1/2}$ content 27.6\% for a separation energy of $0.375$~MeV, in better agreement with our result.  A more elaborated three-body model, including core tensor and pairing correlations \cite{Kik13}, predicted 44\% and 46.9\% of $2s_{1/2}$ and $1p_{1/2}$ admixtures, respectively. Furthermore, the calculation performed by Barranco and co-workers \cite{Barranco01}, describing the neutrons in a mean-field potential supplemented by a pairing interaction and the coupling to the core collective excitations, resulted in 40\% and 58\% for $2s_{1/2}$ and $1p_{1/2}$.
For the present calculations to reproduce the (p,d) experimental data, our model requires a smaller $p$-wave content that those in Refs.~\cite{Barranco01,Kik13}, thus providing a larger s-wave contribution. Other theoretical works predict different numbers for the $s$-wave and $p$-wave admixture \cite{Garrido99,Garrido03}.  Clearly, this disparity of values calls for further experimental and theoretical work. 

It can be argued that the DWBA framework may be overly simplistic to accurately describe the reaction process, since the low beam energy favours higher-order processes, such as initial- and final-state interactions. These effects are not taken into account in the present formalism, which might hinder the effect of the $^{11}$Li structure on the angular distribution. However, the agreement of the calculations with the experimental data is rather satisfactory, which could be an indication that these effects are not crucial in the transfer process. Moreover, some of the configurations resulting from these higher-order processes could be included as part of the non-resonant continuum responsible for the bump at high $n$-$^9$Li energies that appears in Fig.~2 of Ref.~\cite{Sanetullaev16}. This background was already subtracted from the  experimental angular distribution shown in  Fig.~\ref{fig:dsdw}, thus reducing the degree of unsuitability of the DWBA formalism.


\section{\label{sec:sum} Summary and conclusions}
To summarize, we have studied the $^{11}$Li(p,d)$^{10}$Li reaction using the DWBA framework, computing the required 
$\langle ^{11}{\rm Li} | ^{10}{\rm Li} \rangle$ overlap functions from a three-body model calculation of the $^{11}$Li ground state. Our model provides absolute cross sections, in contrast to previous approaches. The comparison of our calculations with the angular distribution reported in Ref.~\cite{Sanetullaev16} confirms the $\ell=1$ dominance of this distribution. While different three-body models are found to explain equally well the shape of the angular distribution, we find a strong sensitivity on the $p_{1/2}$-wave content, with the best agreement provided by a model with $\sim$31\% for this configuration. The coupling with the intrinsic spin of the $^9$Li core does not affect the overall shape of the angular distribution, but changes the height and width of the relative energy spectra. Experimental data on other reactions involving $^{10,11}$Li are to be tested against the same structure input to find better constraints on their properties.

The methodology employed in the present case could be useful to analyze similar data involving other Borromean nuclei, such as $^{6}$He, $^{14}$Be or $^{22}$C. Moreover, the calculated three-body overlaps can be also used in other reactions, such as quasifree $(p,pn)$ breakup processes or $(d,p)$ reactions. Calculations of this kind are in progress and will be presented elsewhere.

\section*{Acknowledgements}
We are grateful to Dr.~R.~Kanungo for providing us the experimental data in tabular form and for useful discussions on their interpretation. This work has received funding from the Spanish Ministerio de Economía y Competitividad under Project No.~FIS2014-53448-C2-1-P and by the European Union Horizon 2020 research and innovation program under Grant Agreement No.~654002. M.G.-R.~acknowledges support from the Spanish Ministerio de Educaci\'on, Cultura y Deporte, Research Grant No.~FPU13/04109.

\section*{References}

\bibliography{mybibfile}

\end{document}